      [1999/12/01 v1.4c Il Nuovo Cimento]
\documentclass{cimento}

\title{Top differential cross section measurements (Tevatron)}
\author{Andreas~W.~Jung\from{ins:x}~~(for the \dzero~collaboration)}
\instlist{\inst{ins:x} Fermi National Accelerator Laboratory (Fermilab), USA \\
        E-mail: ajung@fnal.gov \\ FERMILAB-CONF-11-677-PPD}

\PACSes{\PACSit{13,14}
\PACSit{13,14}}

\usepackage{amssymb,amsmath,array}
\usepackage{wrapfig,rotating}
\usepackage{xspace}

\begin{document}

\newcommand{\dzero}     {D\O\xspace}
\newcommand{\wplus}     {$W+$jets\xspace}
\newcommand{\zplus}     {$Z+$jets\xspace}
\newcommand{\muplus}    {$\mu +$jets\xspace}
\newcommand{\eplus}     {$e +$jets\xspace}
\newcommand{\ljets}     {$l +$jets\xspace}
\newcommand{\ttbar}     {$t\bar{t}$\xspace}
\newcommand{\met}       {$\not\!\!E_T$\xspace}

\maketitle


\begin{abstract}
Differential cross sections in the top quark sector measured at the Fermilab Tevatron collider are presented. CDF used $2.7~\mathrm{fb^{-1}}$ of data and measured the differential cross section as a function of the invariant mass of the \ttbar system. The measurement shows good agreement with the standard model and furthermore is used to derive limits on the ratio $\kappa /M_{Pl}$ for gravitons which decay to top quarks in the Randall-Sundrum model. \dzero used $1.0~\mathrm{fb^{-1}}$ of data to measure the differential cross section as a function of the transverse momentum of the top quark. The measurement shows a good agreement to the higher order perturbative QCD prediction and various predictions based on various Monte-Carlo generators.
\end{abstract}

\section{Introduction}
The $top$ quark is the heaviest known elementary particle and was discovered at the Tevatron $p\bar{p}$ collider in
1995 by the CDF and \dzero collaboration \cite{top_disc1, top_disc2} at a mass of around $175~\mathrm{GeV}$. The production is dominated by the $q\bar{q}$ annihilation process with 85\% as opposed to gluon-gluon fusion which contributes only 15\%. Both measurements presented here are performed using the \ljets~channel, where one of the $W$ bosons (stemming from the decay of the $top$ quarks) decays leptonically. The other $W$ boson decays hadronically. The \ljets~channel is a good compromise between signal and background contribution whilst having high event statistics. The branching fraction for top quarks decaying into $Wb$ is almost 100\%. Jets containing a beauty quark are identified by means of a neural network (NN) build by the combination of variables describing the properties of secondary vertices and of tracks with large impact parameters relative to the primary vertex.

\section{Measurement of the transverse momentum distribution of the top quarks}
\begin{figure}[ht]
    \centerline{\includegraphics[width=0.975\columnwidth]{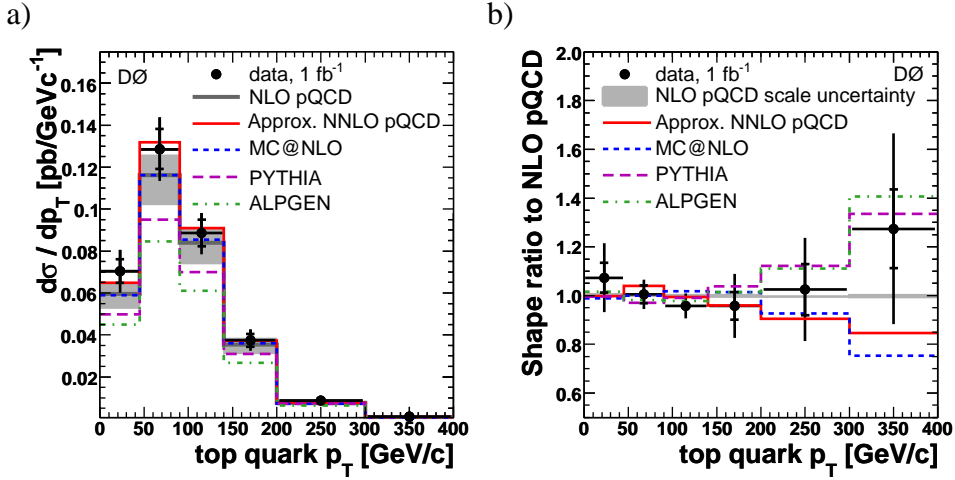}}
  \caption{\label{fig:toppt_xsec} a) Differential cross section data (points) as a function of top-quark $p_T$ (two entries per event) \cite{d0_toppt} compared with expectations from NLO pQCD (solid lines), from an approximate NNLO pQCD calculation, and for several event generators (dashed and dotdashed lines). The gray band reflects uncertainties on the pQCD scale and parton distribution functions. Inner error bars represent the statistical uncertainty, whereas the outer one is statistical and systematic added in quadrature. b) shows the ratio of $(1/\sigma) d\sigma/dp_T$ relative to NLO pQCD for an approximate NNLO pQCD calculation and of predictions for several event generators.}
 \end{figure}

The measurement of the transverse momentum distribution of the top quarks \cite{d0_toppt} selects events with an isolated lepton with a transverse momentum $p_T$ of at least $20~\mathrm{GeV}$ and a pseudo-rapidity of $|\eta| < 1.1$ (\eplus) or $|\eta| < 2.0$ (\muplus). A cut
on the missing transverse energy (\met) of $20~\mathrm{GeV}$ is applied. Furthermore at least four jets are required 
with $p_T > 20~\mathrm{GeV}$ and $|\eta| < 2.5$, an additional cut of $p_T > 40~\mathrm{GeV}$ is applied for the
leading jet. Finally at least one jet needs to be identified as a $b$-jet. For the reconstruction of the event kinematics
additional constraints are used: the masses of the two $W$ bosons are constrained to $80.4~\mathrm{GeV}$. Furthermore the masses of the two reconstructed top quarks are assumed to be equal. All possible permutations of objects are considered where the final solution is the one with the smallest $\chi^2$. \\
 \begin{figure}[ht]
     \centerline{\includegraphics[width=1.\columnwidth]{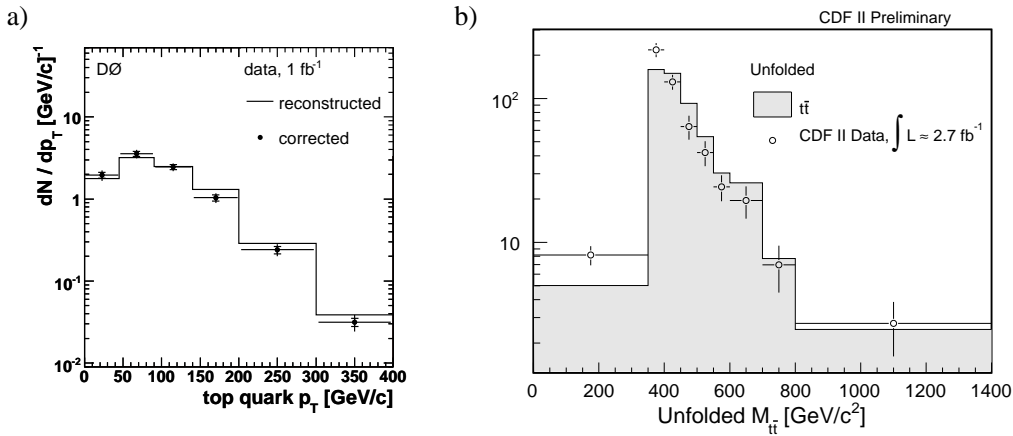}}
   \caption{\label{fig:toppt_unfold} a) compares the background-subtracted reconstructed top-quark $p_T$ distribution \cite{d0_toppt} with the one corrected for the effects of finite experimental resolution (two entries per event). Inner error bars represent the statistical uncertainty, whereas the outer one is statistical and systematic added in quadrature. b) shows the unfolded invariant mass distribution of the \ttbar system \cite{cdf_mtt} compared to signal \ttbar MC.}
  \end{figure}
Figure \ref{fig:toppt_unfold}a) shows the background-subtracted reconstructed top-quark $p_T$ distribution compared to the one corrected for finite experimental resolution. The latter is derived by using regularized matrix unfolding. Figure \ref{fig:toppt_unfold}b) compares the unfolded invariant mass distribution of the \ttbar system to the expectation using \ttbar signal MC. The correction for finite detector resolution is again done using regularized unfolding.\\
Figure \ref{fig:toppt_xsec}a) shows the differential cross section as a function of top-quark $p_T$, where the leptonic and hadronic decay of the $W$ boson to the top-quark cross section are combined. All predictions use the proton parton density function (PDF) CTEQ61 with the scale set to $\mu_{r} = \mu_{f} = m_t$ ($m_t = 170~\mathrm{GeV}$) except for the approximate NNLO perturbative QCD (pQCD) prediction which uses the MSTW08 PDF. The normalization is nicely described by pQCD in (N)NLO, however there is an offset for PYTHIA and ALPGEN in normalization. Figure \ref{fig:toppt_xsec}b) shows that the shape is reasonable described by all predictions. The inclusive total cross section for \ttbar production is measured to $\sigma =  8.31 \pm 1.28 (\mathrm{stat.})~\mathrm{pb}$ and in good agreement with the latest theoretical predictions of $\sigma =  6.41 \pm^{0.51}_{0.42} ~\mathrm{pb}$ \cite{topxsec_theo1} and $\sigma =  7.46\pm^{0.48}_{0.67} ~\mathrm{pb}$ \cite{topxsec_theo2}.

\section{Measurement of the invariant mass distribution of the $t\bar{t}$ system}
\begin{figure}[ht]
    \centerline{\includegraphics[width=0.975\columnwidth]{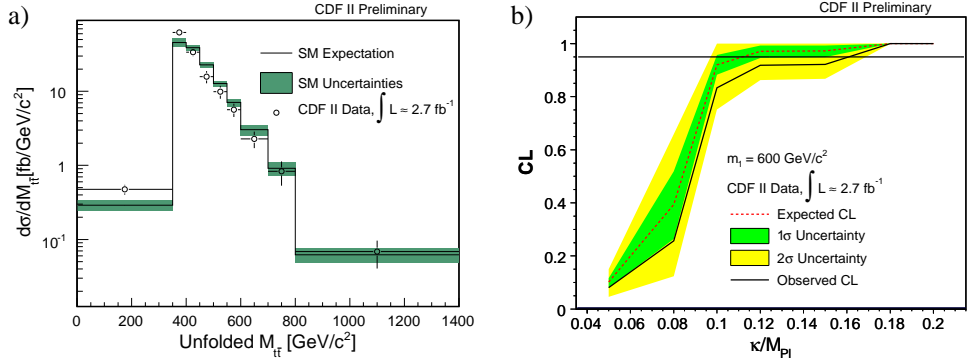}}
  \caption{\label{fig:mtt_xsec} a) shows the differential \ttbar cross section (circles) as a function of $M_{t\bar{t}}$ \cite{cdf_mtt} compared to the standard model expectation (line). The SM uncertainty (green band) reflects all systematic uncertainties, except for the luminosity uncertainty in each bin. b) shows limits on the ratio $\kappa /M_{Pl}$ for gravitons which decay to top quarks in the Randall-Sundrum model, where the mass of the first resonance is fixed at 600 GeV. Values of $\kappa /M_{Pl} > 0.16$ are excluded at the 95\% confidence level.}
 \end{figure}

This measurement of the invariant mass distribution of the $t\bar{t}$ system $M_{t\bar{t}}$ \cite{cdf_mtt} selects events with an isolated lepton with a $p_T$ of at least $20~\mathrm{GeV}$ and a pseudo-rapidity of $|\eta| < 1.1$. A cut on the missing transverse energy of $20~\mathrm{GeV}$ is applied. Furthermore at least four jets are required with $p_T > 20~\mathrm{GeV}$ and $|\eta| < 2.0$. Finally at least one jet needs to be identified as a $b$-jet. The hadronic $W$ decay is used to constrain the Jet Energy Scale (JES). $M_{t\bar{t}}$ is reconstructed by using the four-vectors of the b-tagged jet and the three remaining leading jets in the event, the lepton and the transverse components of the neutrino momentum, given by \met. \\
Figure \ref{fig:mtt_xsec}a) shows the differential \ttbar cross section as a function of $M_{t\bar{t}}$ compared to the standard model expectation using the proton PDF CTEQ5L PDF with a top mass of $175~\mathrm{GeV}$. The SM uncertainty reflects all systematic uncertainties, except for the luminosity uncertainty in each bin. Especially the tail of $M_{t\bar{t}}$ is sensitive to broad enhancements as well as to narrow resonances, which is why the agreement between data and SM expectation has been evaluated. There is no indication of beyond standard model contributions to the differential cross section. The analysis also measured the inclusive total cross section for \ttbar production to: $\sigma =  6.9 \pm 1.0 (\mathrm{stat.+JES})~\mathrm{pb}$, which is in good agreement with latest theoretical predictions \cite{topxsec_theo1,topxsec_theo2} as well as with the \dzero result. Furthermore the distribution has been used to derive a limit on gravitons which decay to top quarks in the Randall-Sundrum model. The mass of the first resonance is fixed to 600 GeV and gravitons are modeled using MadEvent plus Pythia. Figure \ref{fig:mtt_xsec}b) shows the derived limits, values of $\kappa /M_{Pl} > 0.16$ are excluded at the 95\% confidence level.

\section{Conclusion}
Two differential cross section measurements have been presented. The cross section as a function of the transverse momentum of the top quark by \dzero \cite{d0_toppt} and as a function of the invariant mass of the \ttbar system by CDF \cite{cdf_mtt}. Both presented results are consistent with the standard model cross section predictions. The final Tevatron data sample has $5-10$ times the presented statistics allowing for more precise measurements in the near future being one of the legacy measurements of the Tevatron.

\end{document}